\documentclass[twocolumn,superscriptaddress,floatfix,preprintnumbers]{revtex4}
\usepackage{graphics}
\usepackage{amsmath}
\usepackage{amsfonts}
\usepackage{amssymb}
\usepackage{graphicx}

\begin{document}
\title{Fully-automated optimization of grating couplers}

\author{Logan Su}
\author{Rahul Trivedi}
\author{Neil V. Sapra}
\author{Alexander Y. Piggott}
\affiliation{Ginzton Laboratory, Stanford University, Stanford, California 94305, USA}
\author{Dries Vercruysse}
\affiliation{Ginzton Laboratory, Stanford University, Stanford, California 94305, USA}
\affiliation{Department of Physics, KU Leuven, Celestijnenlaan 200 D, B-3001 Leuven, Belgium}
\author{Jelena Vu\v{c}kovi\'{c}}
\affiliation{Ginzton Laboratory, Stanford University, Stanford, California 94305, USA}

\begin{abstract}
We present a gradient-based algorithm to design general 1D grating couplers without any human input from start to finish, including a choice of initial condition. We show that we can reliably design efficient couplers to have multiple functionalities in different geometries, including conventional couplers for single-polarization and single-wavelength operation, polarization-insensitive couplers, and wavelength-demultiplexing couplers. In particular, we design a fiber-to-chip blazed grating with under 0.2 dB insertion loss that requires a single etch to fabricate and no back-reflector.
\end{abstract}

\maketitle

\section{Introduction}
Edge couplers and grating couplers are the primary interfaces used between integrated photonic circuits and optical fibers. Grating couplers are attractive because they are typically easier to fabricate, are flexible in their placement, and enable wafer-scale testing. However, grating couplers tend to have lower coupling efficiencies \cite{chrostowski2015silicon}.

The simplest coupler consists of a uniform grating; however, the maximum efficiency of such a grating is limited \cite{orobtchouk2000high}. For instance, the coupling loss is more than 2.6 dB  for 220 nm thick silicon-on-insulator (SOI) architecture \cite{bozzola2015optimising}. Higher coupling efficiencies can be achieved in a wide variety of ways, including nonuniform gratings \cite{taillaert2004compact, chen2010apodized}, bottom reflectors \cite{van2007compact, zaoui2012cost}, multiple layers \cite{wang2004compact, sacher2014wide, michaels2017inverse}, multiple etch depths \cite{li2013cmos, chen2017dual}, silicon overlays \cite{roelkens2006high, vermeulen2010high}, blazed gratings \cite{matsumoto1992analysis, wang2005embedded, yang2011high}, and unconventional geometries \cite{na2011efficient, sanchez2016broadband}.

A direct consequence of the diversity of grating geometries is that grating couplers need to be optimized to the specific geometry and desired functionalities. Extensive literature exists on optimizing grating couplers \cite{taillaert2004compact, roelkens2006high, andkjaer2010topology, covey2013efficient, zaoui2014bridging, zhong2014focusing, wohlfeil2014optimization, shi2014high, bozzola2015optimising, michaels2017inverse}, but these optimizations often rely on starting with a standard design \cite{bozzola2015optimising, michaels2017inverse}. For conventional geometries and designs, there is well-known analytical theory to suggest an appropriate starting condition. However, for unconventional geometries or devices with multiple functionalities (e.g. wavelength demultiplexing), analytical theory becomes more challenging. In addition, many grating optimization techniques rely on parameter sweeps, random perturbations, or population-based metaheuristic algorithms, such as genetic algorithms and particle swarm optimization, all of which can be time-consuming to perform.

In contrast, gradient-based methods have been promising in designing a wide variety of nanophotonic structures owing to their ability to explore a larger design space. This is possible because gradient-based methods requires only one forward simulation to calculate the fields and one ``backward'' simulation to calculate the gradient by using the adjoint method (see supplementary material in \cite{piggott2017fabrication}).

The optimization landscape of discrete, fabricable structures is highly non-convex and difficult to navigate. Consequently, gradient-based optimization over this space requires finding a suitable initial condition. To automate this process, the {\it discrete optimization stage} can be preceded with a simpler optimization problem whereby the permittivity distribution is allowed to vary continuously between that of the cladding and that of the device. A properly chosen {\it discretization} for converting the resulting structure of this {\it continuous optimization stage} into a starting structure for the discrete stage is critical for achieving efficient devices.

In this work, we present such a two-stage gradient-based optimization algorithm for 1D uniform grating couplers. We compare three different discretization methods and show that our choice of discretization procedure can reliably design efficient gratings using completely random initial conditions, thus fully automating the design process. To illustrate the flexibility of our method, we design a wide class of fiber-to-chip grating couplers, including polarization-insensitive couplers, wavelength-demultiplexing couplers, and highly efficient single-wavelength couplers. Notably, we design a blazed grating coupler with under 0.2 dB loss requiring only a single etch step to fabricate. 

\section{Optimization method}

\subsection{Nanophotonic Inverse Design}
The nanophotonic inverse design problem is given by
\begin{equation}
\label{eqn:invdes}
\begin{aligned}
& \underset{p, E_1, E_2, \dots, E_m}{\text{minimize}}
& & \sum_i f_i(E_i) \\
& \text{subject to}
& & \nabla \times \frac{1}{\mu_0} \nabla \times E_i - \omega_i^2 \epsilon(p)E_i   = -i\omega_i J_i, \\
& & & i = 1, 2, \dots, m
\end{aligned}
\end{equation}
where $m$ is the number of modes, $E_i$ is the electric field at $\omega_i$, $J_i$ is electric field source, $p$ is vector that parametrizes the structure, and $f_i$ is the objective function. $f_i$ is either equal to $f_i^M$ to optimize power in a waveguide mode or $f_i^P$ to optimize power across a plane. Specifically, $f_i^M$ is defined by
\begin{align}
	f_i^M(E_i) = I_+\left(\left|\mathcal{L}(E_i)\right| - \alpha_{i}\right) + I_-\left(\left|\mathcal{L}(E_i)\right| - \ \beta_{i}\right)
\end{align} 
where $\mathcal{L}(E_i)$ is the overlap integral with the waveguide mode as defined in the supplementary material of \cite{ piggott2017fabrication}; $I_+$ and $I_-$ are continuous relaxations of indicator functions as defined in the supplementary material of \cite{piggott2017fabrication}; and $\alpha_i = 1$ and $\beta_i = 0.99$ when maximizing the power and $\alpha_i = 0.01$ and $\beta_i = 0$ when minimizing power. $f_i^P$ is defined by
\begin{align}
	f_i^P(E_i) = \int \Re\left[E_i \times H_i^* \right] dS
\end{align}
where $H_i$ is the magnetic field and the integration is performed over the desired plane.

In our simulation domain, we specify a rectangular {\it design region} within which the grating resides. The permittivity of the design region is determined by a {\it parametrization vector} $p$, while the permittivity distribution outside the design region is fixed.

Our optimization algorithm is broken down into two stages: continuous and discrete. In each stage, the optimization problem described in Equation \ref{eqn:invdes} is solved with different parametrizations of the structure. A discretization process converts the optimized structure from the continuous stage into the initial structure for the discrete stage.

In the continuous optimization stage, the design region is divided into equally spaced pixels, with each element of $p$ representing each pixel. Each pixel takes on a value between 0 and 1, where 0 represents the cladding and 1 represents the device. We optimize using the second-order L-BFGS-B algorithm \cite{byrd1995limited} for a fixed number of iterations or until convergence, whichever occurs first. 

In the discrete optimization stage, the structure is parametrized by the location of the edges of the grating grooves, with each element of $p$ representing a single edge. Under this parametrization, the structure represents a discrete, fabricable device. Feature size constraints are implemented by constraining the distance between the edge locations to be at least the minimum feature size. In order to handle these constraints, we optimize using another gradient-based algorithm, SLSQP \cite{kraft1988software}.

We simulate the grating couplers using the finite-difference frequency domain (FDFD) method \cite{Shin:12, Shin:13}. All simulations are performed with a spatial discretization of 20 nm. The simulation region is surrounded by perfectly matched layers (PMLs) on all four sides. We model the fiber mode as a Gaussian beam with a waist $\sigma_w$ and use an input current source of the form $\exp(-x^2/\sigma_w^2)$.

Since typical grating coupler sizes are on the order of 10 $\mu$m, optimizing full 3D grating couplers is computationally expensive. Instead, we simulate the gratings in 2D. Fortunately, the difference in performance between 2D and 3D simulations is often negligible; for instance, when coupling to 12 $\mu$m strip waveguide at 1550 nm, the 3D structure has an efficiency roughly 97\% of that of the 2D device \cite{taillaert2004compact}. Nonetheless, we emphasize that our reported efficiencies are for 2D coupling efficiencies; the exact achieved efficiency in 3D depends on the length of the coupler in the third dimension.

Roughly 700 simulations were required per mode of the optimization problem, and the total simulation time was approximately 2$m$ hours on a single 6-core Intel Core i7 machine where $m$ is the number of modes.

\subsection{Discretization}
\label{sec:discretization}
Since discrete optimization is inherently harder than continuous optimization and the number of grating edges is fixed in discrete stage, it is imperative to start with a good initial condition for the discrete stage to achieve structures with the high efficiency. In this section, we discuss three possible discretization methods.

One simple way of converting a continuous structure into a discrete one is via simple thresholding whereby pixels whose values are greater than $\frac{1}{2}$ are set to 1 and pixels less than $\frac{1}{2}$ are set to 0. However, this method results in many closely-spaced grating edges, which performs poorly when feature constraints force the the edge locations to spread out. Variations on thresholding, including post-processing the structure to remove closely-spaced edges, also do not perform well.

Rather than developing a hand-crafted heuristic algorithm to perform discretization and post-processing, the discretization task can be as an optimization problem. Intuitively, a good initial condition is one that is similar to the optimized continuous structure. This sentiment can be formalized through an optimization problem, which we will call the {\it least-squares discretization (L2D)}:
  \begin{equation}
  \label{eqn:opt1}
 \begin{aligned}
 & \underset{p,n}{\text{minimize}}
 & & ||R(p) - q||_2 \\
 & \text{subject to}
 & & p_{i+1} \geq p_i + d \\
 & & & p_1 \geq 0 \\
 & & & p_n \leq L
 \end{aligned}
 \end{equation}
 where $p \in \mathbb{R}^n$ is a vector of edge locations, $q \in \mathbb{R}^m$ is the parametrization of the optimized continuous device, $d$ is the minimum feature size in terms of pixels (fractional pixels are allowed), $L$ is the number of pixels (i.e. design length), and $R: \mathbb{R}^n \rightarrow \mathbb{R}^m$ is a function that takes a vector of edge locations and renders it onto the same grid of pixels as in the continuous optimization. 
 
\begin{figure*}[htb]
	\centering
	\includegraphics[scale=0.9]{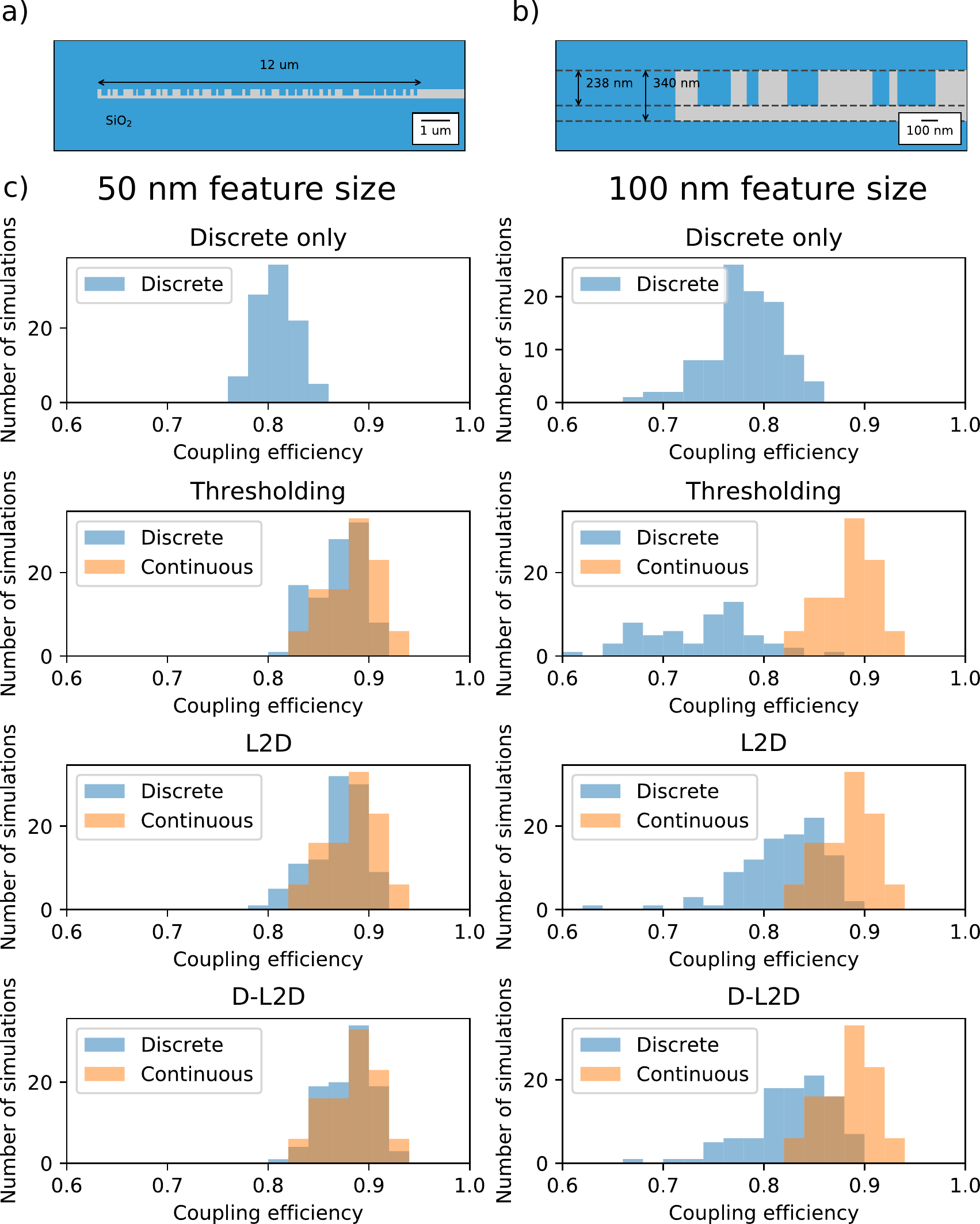}
	\caption{ a) Schematic of a typical optimized grating used in the study. b) Close-up schematic of the device. c) Distribution of efficiencies for 100 optimization runs for different discretization procedures for 50 nm and 100 nm feature sizes. Each optimization run ran for a maximum of 400 iterations (100 for continuous stage and 300 for discrete stage). The orange shows the distribution of efficiencies achieved at the end of the continuous stage while the blue shows the distribution at the end of the discrete stage. Notice that forgoing the continuous optimization step leads to worse-performing devices.}
	\label{fig:single-page}
\end{figure*}

On occasion, we observed L2D produces a discrete device with features, corresponding to regions of weakly-modulated permittivities in the continuous stage. To mitigate this, the optimized continuous structure $q$ is first deconvolved using the optimization problem
\begin{equation}
\begin{aligned}
& \underset{q'}{\text{minimize}}
& & ||Aq' - q||_2 \\
& \text{subject to}
& & 0 \leq q'_i \leq 1
\end{aligned}
\end{equation}
where $A$ is the matrix representation of a convolution kernel. Then, the optimal deconvolved structure $(q')^*$ is used as $q$ in Equation \ref{eqn:opt1}. In our optimizations, we have chosen a convolution matrix corresponding to a moving average across 5 elements. We will refer to this approach as the {\it deconvolved least-squares discretization (D-L2D)}.

\subsection{Optimization statistical study}
\label{sec:study}

We ran the optimization 100 times using randomly generated initial conditions and applying each discretization method described in Section \ref{sec:discretization} for coupling 1550 nm into a 340 nm thick waveguide at normal incidence (Figure \ref{fig:single-page}). We have chosen to use an infinite buried oxide (BOX) layer in order to reduce computation time. In addition, we ran the optimization method without the continuous stage. For these ``discrete-only'' optimization runs, we chose to seed the optimization by first choosing a vector uniformly at random and then applying D-L2D to arrive at a discrete starting condition.

The statistical studies show most choices for initial condition result in reasonably efficient devices. Nevertheless, using optimization to perform discretization is superior than thresholding or forgoing the continuous optimization entirely.

Discrete-only optimization performs significantly worse than L2D and D-L2D for 50 nm feature sizes because of the highly non-convex landscape. The better relative performance at 100 nm arises from the fact that the design space becomes smaller as the feature size increases. Consequently, it is more likely to randomly pick a good initial condition for discrete-only optimization.

Thresholding performs similarly to L2D and D-L2D for 50 nm feature sizes because the small feature size means that the grating can have many closely-spaced edges. However, at 100 nm feature sizes, the grating edges can no longer be placed so densely together, so thresholding performs even worse than the discrete-only optimization. We have found that modifications to thresholding only results in a performance between different feature sizes. For example, post-processing thresholding step to remove closely-spaced edges improved the performance at 100 nm feature size at the expense of poorer performance at 50 nm feature size.

In contrast, L2D and D-L2D perform the best in both cases. D-L2D performs slightly better than L2D in all cases, both in terms of the mean efficiency and the maximum efficiency out of all 100 runs. This is intuitive because L2D performs well when the optimized continuous structure appears mostly discrete, and the deconvolution used in D-L2D does not affect substantially continuous structures that are mostly discrete. Therefore, there is little disadvantage in employing D-L2D over L2D. We have also performed a similar study for wavelength-demultiplexing grating couplers and found similar results.

\section{Grating coupler designs}

In this section, we illustrate our optimization method by designing a wide variety of grating couplers, including polarization-insensitive and wavelength-demultiplexing gratings.

\subsection{Single-function grating couplers}
\label{sec:single-wavelength}

Here we focus on gratings for 1550 nm on the 220 nm thick silicon-on-insulator (SOI) platform. Higher efficiencies can be achieved by using thicker waveguides, but we focus on 220 nm because of its role as a common industry standards \cite{chrostowski2015silicon}.

\begin{figure}
	\centering
	\includegraphics[scale=0.7]{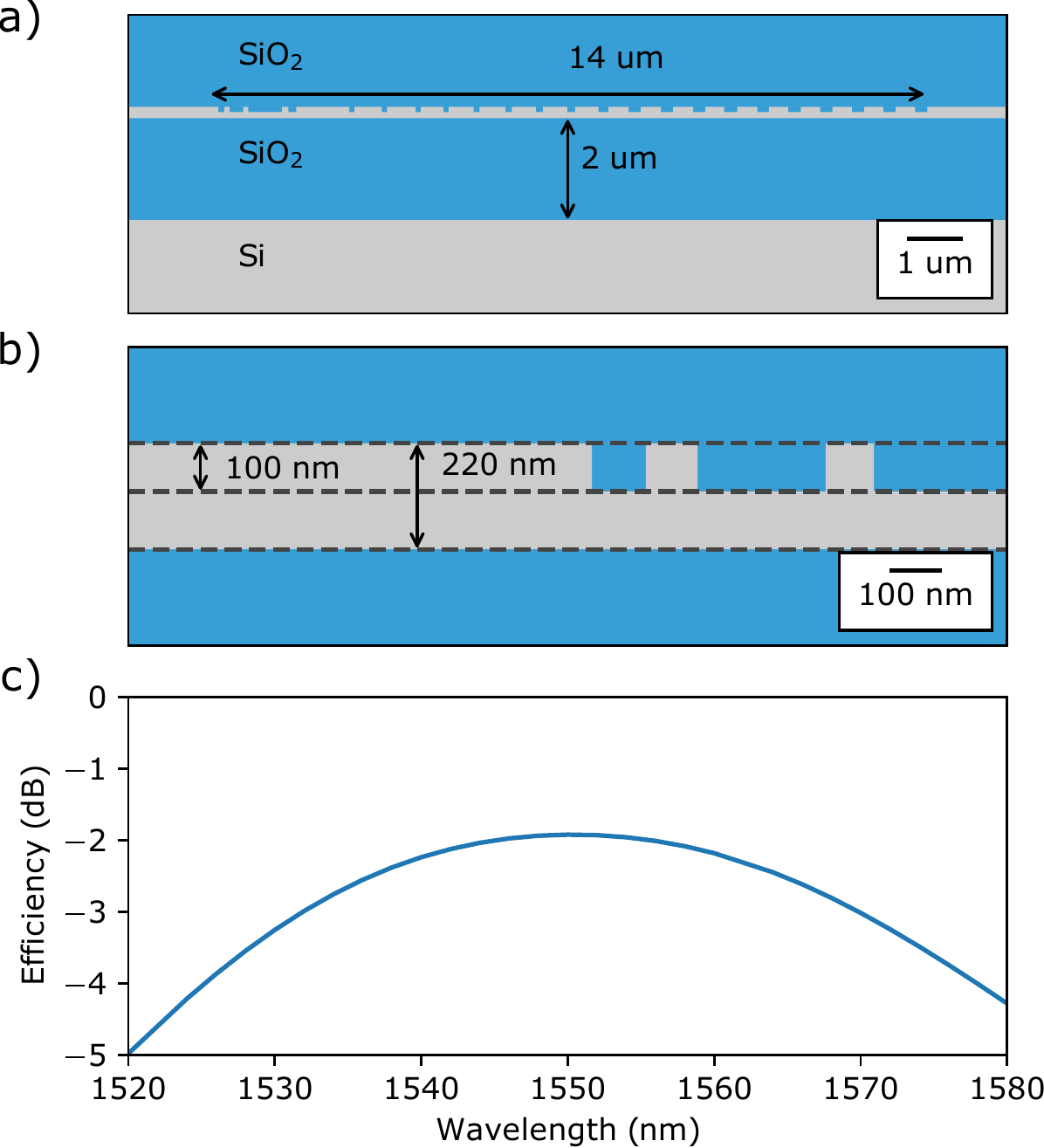}
	\caption{Grating coupler design for 220 nm SOI platform. The angle of incidence is at $10^\circ$ from the normal. a) The grating coupler design. The minimum feature size is 100 nm. b) Close-up schematic of the design. c) Simulated coupling efficiency spectrum. The minimum insertion loss is 1.94 dB, and the 1-dB bandwidth is 34 nm.}
	\label{fig:single-grating}
\end{figure}

Figure \ref{fig:single-grating} shows the design of a conventional grating coupler for 220 nm SOI platform at 10 degree incident angle with 100 nm feature sizes. According to analytical theory, the maximum efficiency is achieved by modulating the etch depth or grating period \cite{bates1993gaussian}. Ideally, the grating period would become smaller and smaller as one approaches the start of the waveguide. This is precisely what occurs in our design. At the beginning of the waveguide, the grating periodicity becomes atypical, but this is a consequence of the desired periodicity falling below the minimum feature size. This apodized design closely resembles previous optimization work in grating couplers \cite{bozzola2015optimising}, and we achieve a similar insertion loss of 1.94 dB (compared to 2.12 dB in \cite{bozzola2015optimising}).

\begin{figure}[htb]
	\centering
	\includegraphics[width=\linewidth]{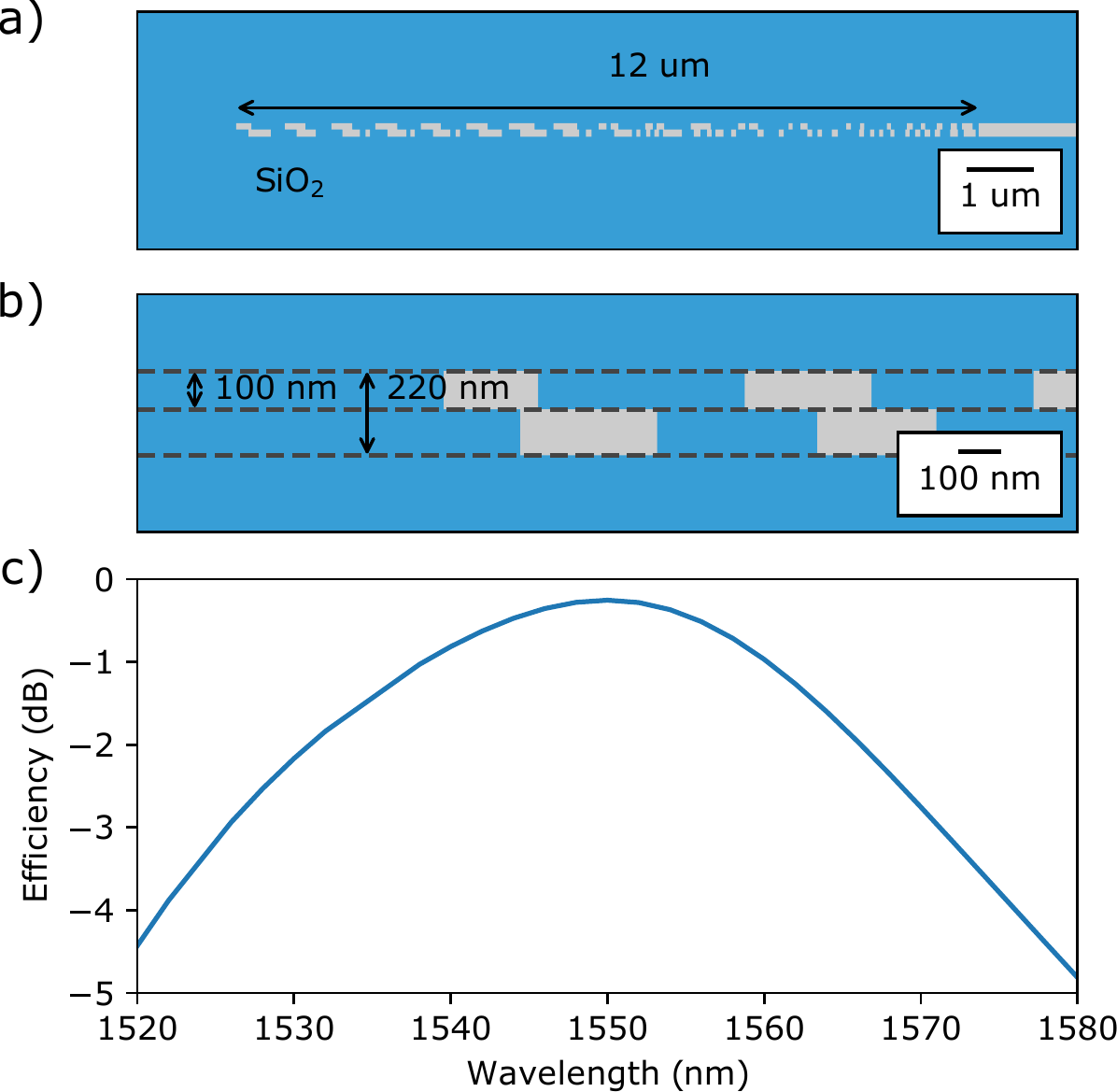}
	\caption{Grating coupler design for 220 nm SOI platform. The device is designed for normal incidence operation. a) The grating coupler design. The minimum feature size is 65 nm. b) Close-up schematic of the design. c) Simulated coupling efficiency spectrum. The minimum insertion loss is 0.25 dB, and the 1-dB bandwidth is 22 nm.}
	\label{fig:yablo}
\end{figure}

In order to achieve higher efficiencies, one can use more complicated geometries. Figure \ref{fig:yablo} shows a two-layer grating structure optimized with our algorithm. This geometry is similar to the one presented in \cite{michaels2017inverse}, and we achieve a similar insertion loss of 0.25 dB (compared to 0.165 dB) with a structure that resembles the one presented in \cite{michaels2017inverse}. However, in \cite{michaels2017inverse}, the initial condition was physically-motivated by considering constructive and destructive interference between the top and bottom layers, whereas our algorithm used a completely random initial condition.

\begin{figure}
	\centering
	\includegraphics[width=\linewidth]{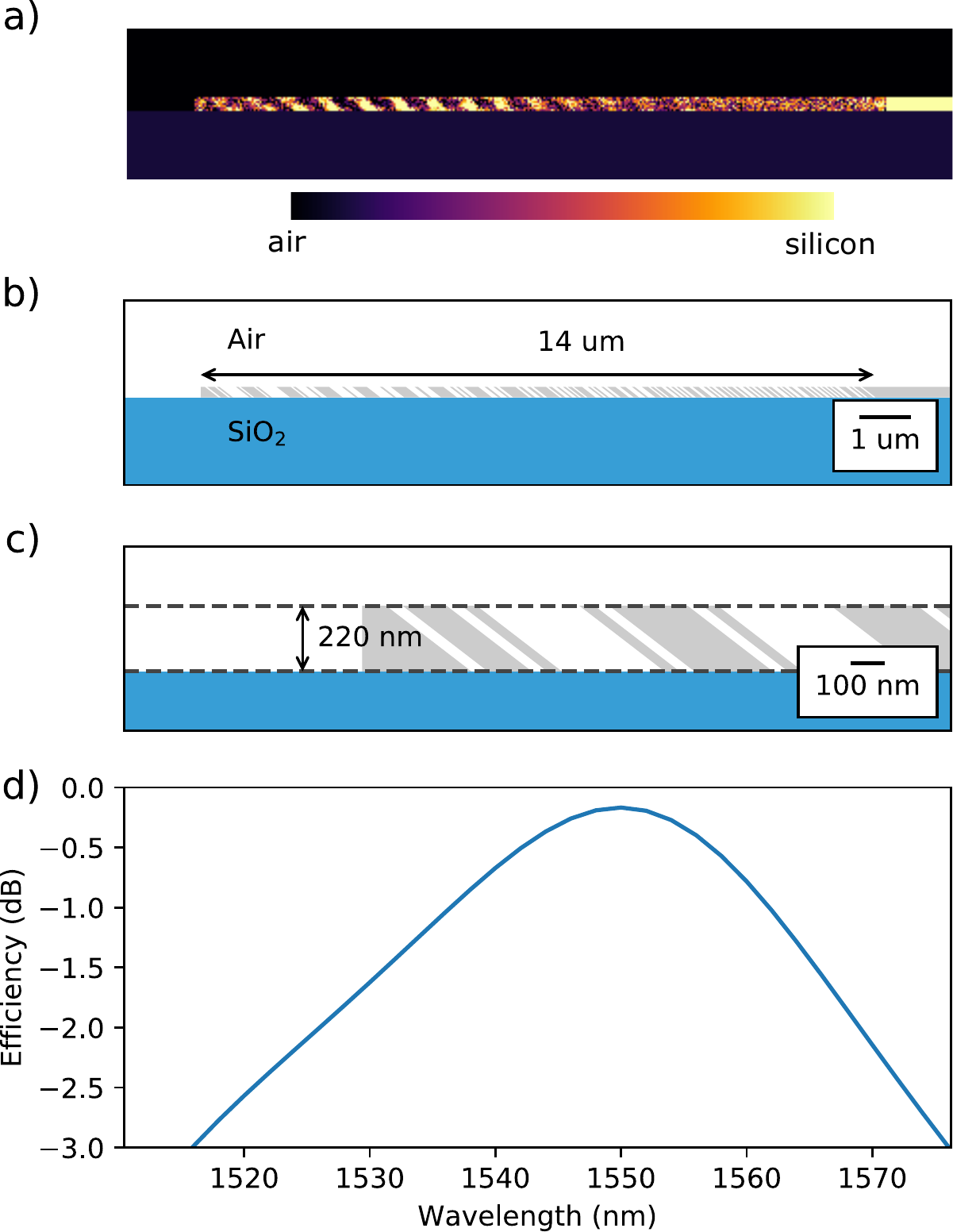}
	\caption{Blazed grating coupler with under 0.2 dB loss. a) Fully continuous design that motivates the blazed gratings. By allowing permittivity distribution in the entire coupler to vary continuously between that of air and silicon, the optimized device has over 99\% coupling efficiency and suggests that blazed gratings would have improved efficiency compared with vertically-etched gratings.  b) The grating coupler design. The device is designed for normal incidence operation. The minimum feature size is 50 nm, and the blazed angle is at 50 degrees from the normal. c) Close-up schematic of the design. d) Simulated coupling efficiency spectrum. The minimum insertion loss is 0.17 dB, and the 1-dB bandwidth is 26 nm. }
	\label{fig:slanted}
\end{figure}

Instead of pre-selecting a particular geometry to optimize, we utilized our method to suggest an optimal geometry for a single-wavelength grating coupler (Figure \ref{fig:slanted}). To achieve this, the structure was parametrized so that all pixels in the design region are allowed to vary continuously. The optimized continuous result clearly suggests that blazed gratings are an optimal geometry, which is consistent with theoretical analysis in \cite{matsumoto1992analysis}. Using this suggestion, we designed a blazed grating with 50 degree slant. In order to handle the slants, the continuous stage optimization was modified so that each value in the parametrization corresponds to a parallelogram pixel with a width and height equal to the spatial discretization. The resulting device has a minimum insertion loss of 0.17 dB with a 26 nm 1-dB bandwidth.

\subsection{Multi-function grating couplers}

In this section, we explore grating couplers that have more than one functionality. Because of the multiple functions, it is more difficult to derive a suitable initial condition analytically, and a fully-automated optimization process becomes particularly useful.

\begin{figure}
	\centering
	\includegraphics[width=\linewidth]{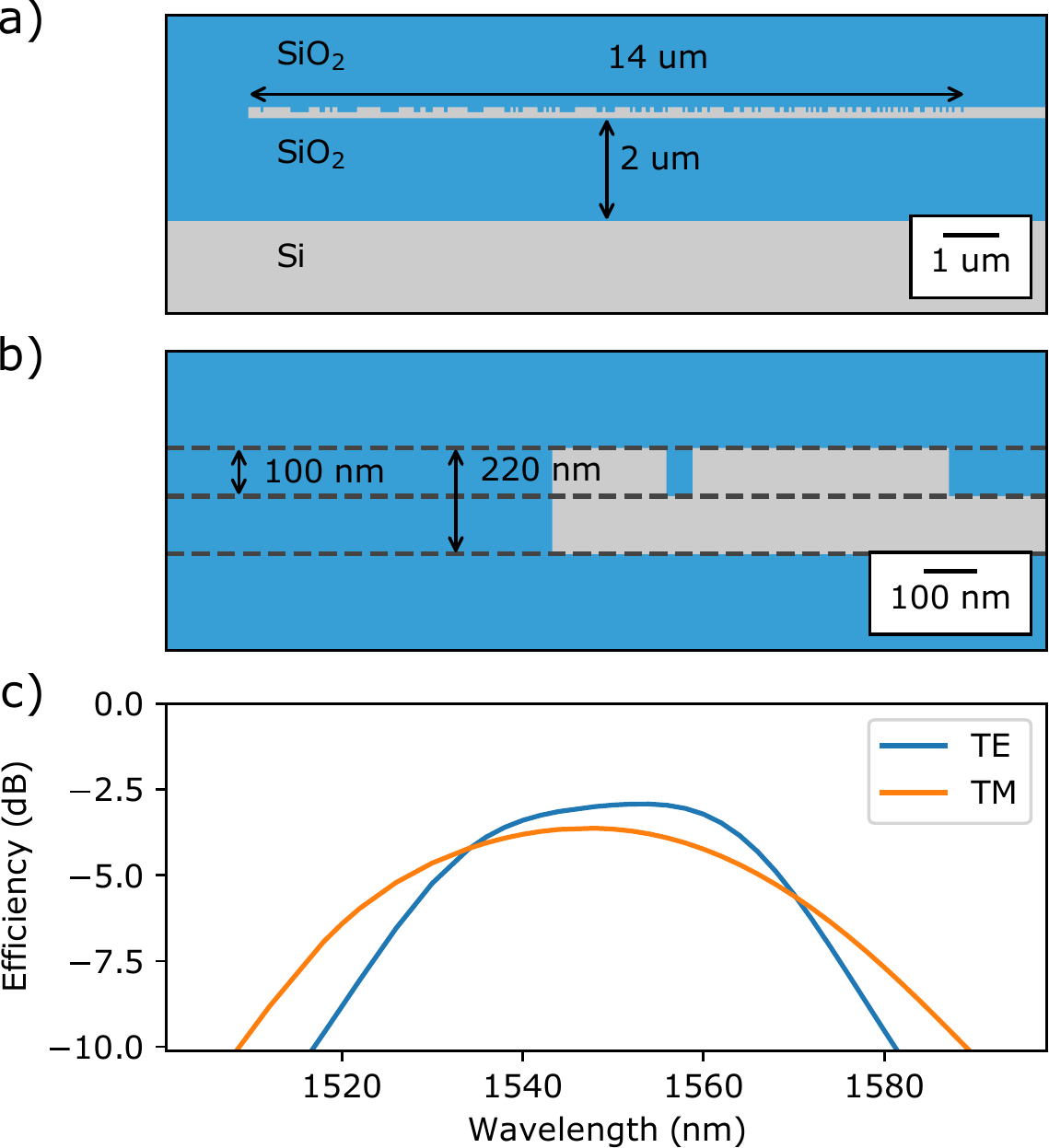}
	\caption{Polarization-insensitive grating coupler design for 220 nm SOI platform. The TE Gaussian mode is coupled to the TE$_0$ mode of the waveguide and the TM Gaussian mode is coupled to the TM$_0$ mode of the waveguide. The device is designed for normal incidence operation. a) The grating coupler design. The minimum feature size is 50 nm. b) Close-up schematic of the design. c) Simulated coupling efficiency spectrum. The minimum insertion loss is 2.9 dB for TE mode and 3.6 dB for TM mode. The 1-dB bandwidth is 28 nm for both modes.}
	\label{fig:single-grating-tm}
\end{figure}

Figure \ref{fig:single-grating-tm} shows a polarization-insensitive grating where the TE Gaussian mode is coupled to the TE$_0$ mode of the waveguide and the TM Gaussian mode is coupled to the TM$_0$ mode of the waveguide. Such gratings are useful because the input fiber is often not polarization-maintaining. Ref. \cite{cheng2014experimental} shows a similar design with 4.3 dB insertion loss for TE and 3.2 dB insertion loss for TM for 340 nm thick waveguides. Our device has an insertion loss of 2.9 dB for TE mode and 3.6 dB for TM mode over a 28 nm bandwidth using 220 nm waveguides.

\begin{figure}
	\centering
	\includegraphics[width=\linewidth]{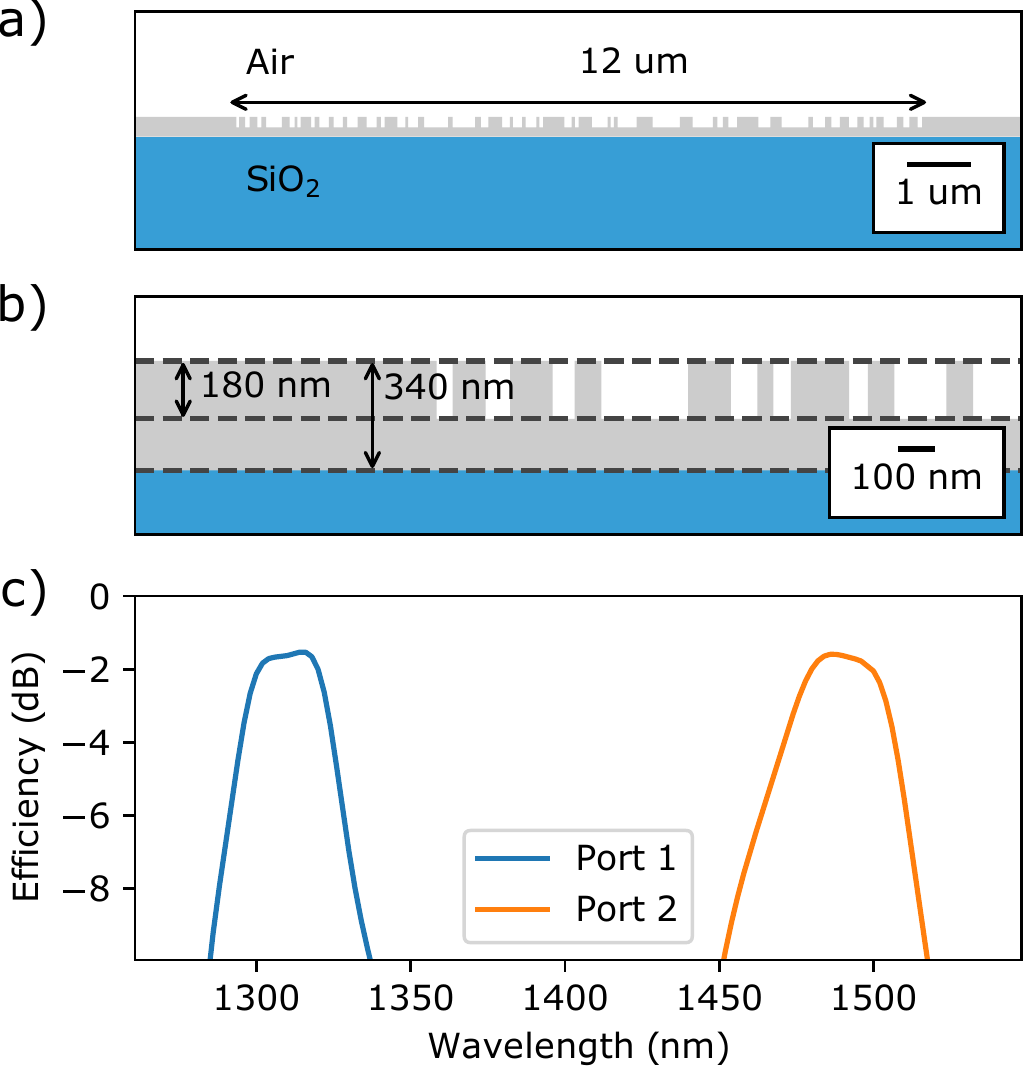}
	\caption{Wavelength-demultiplexing grating coupler that sends 1310 nm into the left waveguide and 1490 nm into the right waveguide. The device is designed for normal incidence operation. a) The grating coupler design. The minimum feature size is 50 nm.  b) Close-up schematic of the design. c) Simulated coupling efficiency spectrum. The minimum insertion loss is 1.5 dB at 1310 nm and 1.6 dB at 1490 nm with over 21 dB crosstalk suppression. The 1-dB bandwidth is 20 nm at 1310 nm and 24 nm at 1490 nm.}
	\label{fig:splitter_wg340}
\end{figure}

Next, we designed a wavelength demultiplexing grating coupler that couple 1310 nm light into the fundamental mode of one waveguide and couple 1490 nm light into the fundamental mode of another waveguide (Figure \ref{fig:splitter_wg340}). Such a grating is useful in wavelength division multiplexing systems where multiple wavelengths are utilized to increase the communication bandwidth. Unlike \cite{roelkens2007silicon} and \cite{streshinsky2013compact}, this grating operates at normal incidence. This is a geometry that we have studied before in \cite{piggott2014} but the design presented there was a focused Gaussian spot rather than one meant for fiber mode. The device presented here has an insertion loss of 1.5 dB at 1310 nm and 1.6 dB at 1490 nm with over 21 dB crosstalk suppression. We note that this was achieved using an infinite BOX layer, which is a more difficult design problem.

\begin{figure*}
	\centering
	\includegraphics[scale=0.8]{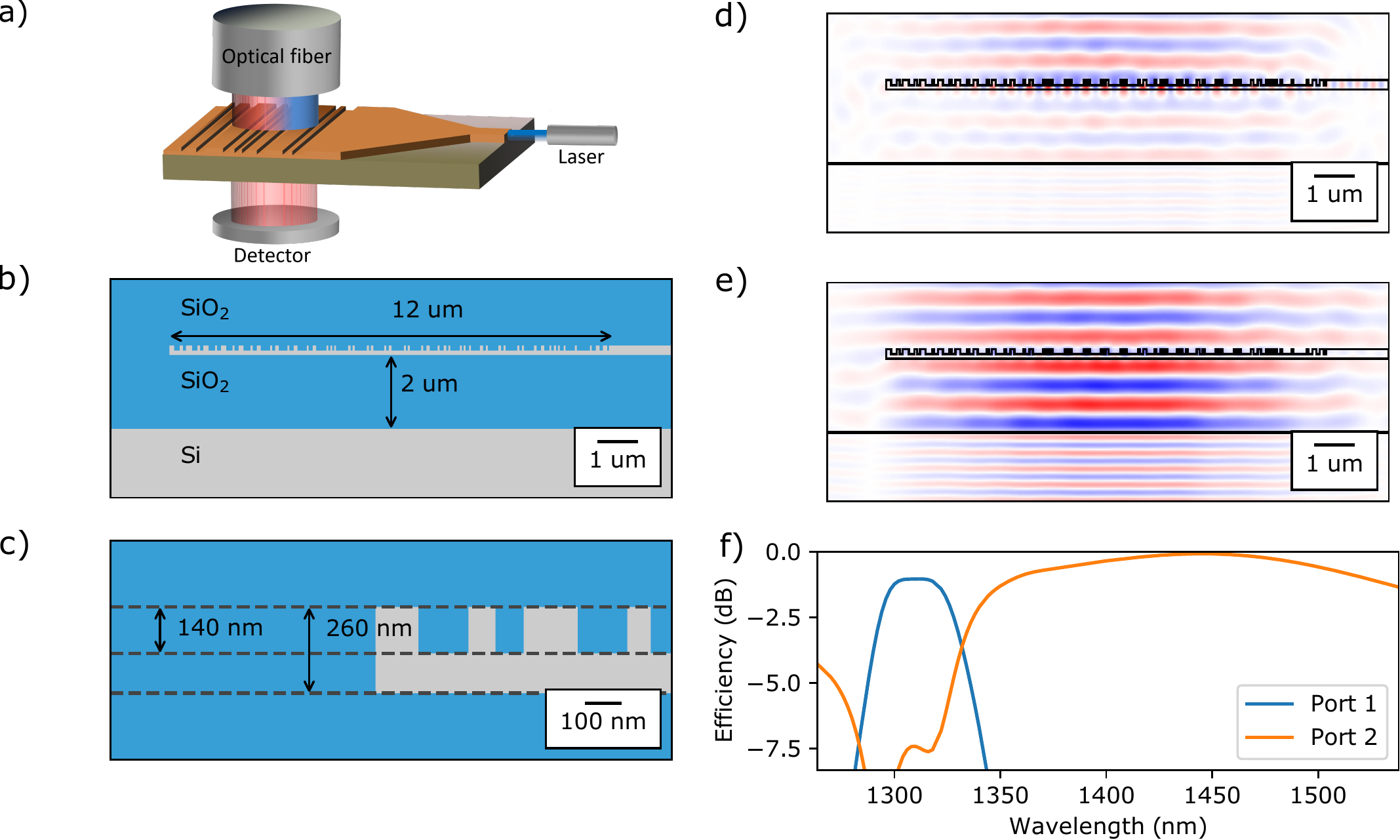}
	\caption{Wavelength-demultiplexing grating coupler that splits 1310 nm and 1490 nm in the pass-through configuration. The device is designed for normal incidence operation. a) The pass-through grating coupler geometry used as an optical transceiver. The transmitting laser (blue) couples through an on-chip waveguide to the fiber whereas the receiving wavelength (red) passes through the grating and is detected with a photodetector. b) The grating coupler design. c) Close-up schematic of the design. The minimum feature size is 50 nm.  d) Real part of the TE mode electric field at 1310 nm. e) Real part of the TE mode electric field at 1490 nm. f) Simulated coupling efficiency spectrum. The minimum insertion loss is 1.0 dB at 1310 nm and 0.08 dB at 1490 nm. The 1-dB bandwidth is 28 nm at 1310 nm.}
	\label{fig:splitter-wg260-pt}
\end{figure*}

To achieve even higher efficiency for wavelength demultiplexing designs, we introduce the {\it pass-through geometry} in which only one of the wavelengths is coupled into an on-chip waveguide whereas another wavelength passes through the grating. A photodetector can then be placed behind the chip to collect the light that passes through the device. This is useful for systems where the pass-through wavelength does not require additional processing. One particular application is in an optical transceiver: The transmitting wavelength would out-couple into an optical fiber through the waveguide whereas the receiving wavelength would pass through the structure and be detected (Figure \ref{fig:splitter-wg260-pt}a). The advantage of this geometry is that the the couplers can be more efficient because high transmission through the grating is easier to achieve. Figure \ref{fig:splitter-wg260-pt} shows a design with 1.0 dB insertion loss at 1310 nm (the on-chip wavelength) and 0.08 dB loss at 1490 nm (the pass-through wavelength).

\section{Conclusion}
We have presented a general gradient-based 1D grating design algorithm that fully automates the design process, enabled by appropriately choosing a least-squares discretization procedure. Using this algorithm, we design efficient couplers in different geometries and with different functionalities, including polarization-insensitive gratings, wavelength-demultiplexing gratings, and a single-wavelength grating coupler with under 0.2 dB insertion loss.

\section{Acknowledgements}
This work was funded by the AFOSR MURI for Aperiodic Silicon Photonics, grant number FA9550-15-1-0335, the Gordon and Betty Moore Foundation, and Huawei.

\end{document}